\documentclass[a4paper,conference]{IEEEtran} 
\ifCLASSINFOpdf
\else
\fi
\usepackage[cmex10]{amsmath}
\usepackage{amssymb,verbatim}
\usepackage{graphicx}

\begin{document}
%
\title{%
Towards Quantum Enigma Cipher III\\
-Communication performance-
}

\author{
\IEEEauthorblockN{Osamu Hirota\\}
\IEEEauthorblockA{
Quantum ICT Research Institute, Tamagawa University\\
6-1-1 Tamagawa-gakuen, Machida, Tokyo 194-8610, Japan\\
{\footnotesize\tt E-mail: hirota@lab.tamagawa.ac.jp} \vspace*{-2.64ex}}
}

\maketitle

\begin{abstract}
 Cloud computing system based on data centers has recently 
attracted considerable attention. In that system, 
all data are communicated via a high speed optical network between 
a customer and data center or between data centers. 
There is a serious threat so called ``Eavesdropper data center business", 
which means the eavesdropper can get all data from the transmission line 
and sell specific data selected by the protocol analyzer to malicious 
people who want to get the secret data. 
So we need to consider cyber attack against Layer-1 (physical layer).
Quantum cryptography has been developed to protect such an attack.
In order to apply such a new security technologies, the communication 
performance is very important as well as its security, because 
the data speed is more than several Gbit/sec.
 This research note III will discuss communication performances 
 of quantum key distribution (QKD) and  quantum enigma cipher, 
 and explaines that QKD based on single photon signals cannot realize 
 appropriate data speed, but quantum enigma cipher can .  
\end{abstract}

%
\IEEEpeerreviewmaketitle
\section{Introduction}
  It has been claimed by many research institutes that quantum key 
distribution allows two remote users to generate a secure random key 
for practical networks. However, in reality, this is not true, because 
communication performance of QKD is extremely poor. The reason is that 
 they have to employ single photon or very weak light as 
the transmission signal.
 QKD researchers would like to simultaneously reach long distances
 and high rates, but it is impossible by the law of physics.
In order to avoid this problem, they employed a theory of 
a rate-loss tradeoff in which they restrict the 
meaning of ``rate" as the term. That is, their rate means
 bit per pulse or mode, not bit per sec that is the most essential 
for communication performance. In general, such an unit has to be 
discussed under the non vanishing situation of signals
 in the communication process.
Let us describe a confusion in the following:\\
\\
$\lceil$To achieve QKD with high rates is demanded if we compete with the 
the conventional infrastructure for fiber based classical communication 
 where the general IT network  sets rate as high as several Gbit/sec
 for distance up to several hundred $km$. So there has been an effort towards 
high-rate QKD.$\rfloor$\\ 

Team of ``high rate" in the above QKD and  Gbit/sec of fiber link has
 no relation in this case.
QKD by single photon signal cannot realize a secure communication for 
 Gbit/sec as data. Despite the fact, QKD group uses ambiguous term.
An example of the typical trick is  a paper entitled
  ``Quantum secured gigabit optical access networks)[1] by 
  A.Shields et al (Toshiba group)." It is clear that 
there is a mismatch between title and contents. 

Recently MIT groups have pointed out that such a description on 
QKD is not fair. They clearly describe the communication performance 
as follows [2,3,4]:\\
\\
$\lceil$QKD would enable full IT security by means of one time pad, but 
 its secret key rate has to equal to the desired communication rate
 (bit/sec). Unfortunately demonstrated key rates fall far short of
  what is needed for widespread use of one time pad. 
  At present, therefore, QKD's principal application is to rekey classical
   cryptosystems, making  overall security that of the classical 
   system. In fact, numerous QKD protocols have demonstrated
 robust extraction of secret key over 200 $km$ with key rates up to 
 the order of 10 bit/sec.$\rfloor$\\

As general users pointed out many times, when we consider 
secure communication systems, we have to take into account 
the following requirements on the encryption system to protect the data of 
gigabit per sec:\\
\\
{\bf{Requirement of specifications:}}\\
 {{(1) Data-speed:}}1 Gbit/sec $\sim$ 100 Gbit/sec\\
 {{(2) Distance:}} 1000 $km$ $\sim$ 10000 $km$ \\
 {{(3) Encryption scheme:}} Symmetric Key Cipher \\
 {{(4) Security:}} Provable security, Secure against Brute force attack
(exhaustive search trial for secret key) by means of computer and also 
 physical devices. \\
 \\
In this note, we will reaffirm that QKD cannot provide such performance, but 
quantum enigma cipher [5] can. 

\section{Basis of communication performance}

\subsection{Single photon communication}
Let us describe the energy loss channel for photon number.
The input-output relation for the photon channel of transmissivity $\kappa$ 
is given as follows:
\begin{equation}
P(n|k)={}_nC_k(\kappa)^k(1-\kappa)^{n-k}
\end{equation}
where $n$ and $k$ are photon number at input and output, respectively.
For the single photon signal, the arrival probability of the photon 
at the channel output is 
\begin{equation}
P(k=1) = \kappa
\end{equation}
Thus, the rate of speed becomes 
\begin{equation}
R_S=\kappa
\end{equation}
The loss property of an optical fiber with low loss is 0.2 dB/km
 for the wavelength 1.55 $\mu$m. This means  $\kappa=0.01$ 
 for the fiber length $100 km$. Before we start QKD protocol, the rate of 
 the channel is already $\kappa=0.01$. So, when the input rate of this channel 
 is 1 Gbit/sec, the output rate is 10 Mbit/sec.

\subsection{Coherent state communication}
Let us consider the transmission of coherent state signal. 
There is an important theorem as follows:\\

${\bf Theorem}$ [6]: Only coherent state can maintain the pure state property against energy loss channel.\\

Thus, the arrival probability of signal itself is 
\begin{eqnarray}
P(pulse)&=&1-|<0|\alpha>_{out}|^2=1-e^{-\kappa|\alpha|^2} \\
&\sim& 1 \quad \quad \kappa|\alpha|^2 >> 1
\end{eqnarray}
where $|\alpha>_{out}$ is the coherent state at the output of 
the channel, and $\alpha$ is the input amplitude of coherent state signal.
In the conventional optical communications, $|\alpha|^2 =10^6$ at 1 Gbit/sec. 
Thus, when the input rate is 1 Gbit/sec, the output is  also 1 Gbit/sec.

\subsection{PPM communication}
The pulse position modulation (PPM) is one of modulation schemes in  
communication systems. A typical structure used in quantum optical 
communication is 
\begin{equation}
|\Psi_i>=|0>_1|0>_2\dots |\alpha>_i|0>_{i+1} \dots |0>_N
\end{equation}
where $i$ and $N$ mean $i$th signal and slot length, respectively.
That is, the information $i$ as the data corresponds to the position of the 
coherent state with non zero amplitude. The total energy per symbol is 
$<n>=|\alpha|^2$. So, the energy per pulse becomes $|\alpha|^2/N$.
Thus, one symbol in this scheme can convey $log_2 N$ bits  under 
the low average energy.
The arrival probability of symbol is 
\begin{equation}
P(symbol)=1-|<0|\alpha>_{out}|^2=1-e^{-\kappa|\alpha|^2}
\end{equation}
It looks very efficient in the sense of the energy constraint when 
we use $N >> 1$.
However, it is not efficient in the sense of frequency bandwidth, 
because this requires $N$ times of the bandwidth of the 
conventional communication systems. So far, many ideas have published 
to improve the bandwidth explosion, keeping a good energy efficiency.
The most famous method is to employ Reed-Solomon code. 
Even though such coding theories can improve the bandwidth problem, still 
there are many serious inefficiencies in the total performance of 
communications.

\section{Communication performance of QKD based on 
single photon or weak light}

\subsection{Rate theory of QKD} 
The rate theory is one of the most important subjects in the theoretical QKD. 
So far, the theoretical analysis of rate for QKD has been dealt with 
the unit of bit per pulse or mode.
Recently, the incomplete bound [7] of the rate has been replaced by 
the rigorous and beautiful analysis [8] as follows:
\begin{equation}
R =\log \frac{1}{1-\kappa} \sim 1.44\kappa \quad bit/pulse
\end{equation}
Since the pulse (photon signal) disappears with 
probability  $\kappa$, the real 
rate (efficiency) for the practical use in the real fiber channel is 
\begin{equation}
R_E =R_S\times R \sim 1.44 \kappa^2
\end{equation}
This means that when the input rate is 1 Gbit/sec,  
the output under the ultimate efficiency is\\ 
\\
\quad \quad (i) 100 Kbit/sec for 100 $km$,\\
\quad \quad (ii) 10 bit/sec for 200 $km$. \\
\\
This fact supports 
the comment of MIT paper [2-4] refered  in the introduction of this paper.
Thus, QKD has a serious defect in the sense of communication performance.
To date, nobody is interested in such an inefficient communication system.
Thus, the author cannot understand how the QKD system can provide the security 
for  ``gigabit access network [1]". The real situation is that QKD experiment 
has been demonstrated in the conventional fiber transmission system 
of gigabit data communication. Thus their title is misleading.
In fact, the headquarter of Toshiba believed that real gigabits data 
has been encrypted by quantum system. 

\subsection{Cryptosystem based on QKD}
(1) AES+QKD
\\
It is clear that the QKD is the most inefficient scheme in modern 
communication systems, so we have to use the QKD
 for providing the secret key of a mathematical encryption such as AES 
 (Advanced Encryption Standard).
 AES has in general the secret key of 256 bits. QKD can provide 256 bits 
 with the delay time of 26 seconds, because the bit rate is 10 bits/sec 
when the communication distance is 200 km.
 Thus we need 26 seconds for one round of refresh key.
 AES operating at 1 Gbit/sec sends 26 Gbits during each 26 seconds 
 based on the previous key. The eavesdropper can get the correct ciphertext 
  of 26 Gbits from the transmission line. This is sufficient to launch 
  the crypto-analysis. In addition, AES can be decrypted by the Brute force 
  attack under the correct cipher text and plaintext of 256 bits.
  This means that the plaintext (data) of 26 Gbits -256 bits can be 
  decrypted by only 256 bits, in principle.\\
  \\
$\quad$ (2) One time pad
  \\
  Let us examine the one time pad system. We assume that \\
  \\
\quad   (i) We want to encrypt 1 Gbits data.\\
\quad  (ii) Transmission systems can send 1 Gbits/sec\\
\\
To encrypt 1 Gbits, we have to store the generated key sequence of 1 Gbits 
 in the Hard-Disk by spending $10^8$ seconds (3 years). 
Then 1 Gbits is sent at 1 second. Again we have to wait $10^8$ 
seconds (3 years) to send the next 1 Gbits.

\section{Communication performance of quantum enigma cipher}
\subsection{Rate of quantum enigma cipher by coherent state}
As discussed in the previous section, the signal does not disappear in 
the coherent state communication with large energy like typical classical
 communication system. Since the quantum enigma cipher (QEC) can employ 
 such large energy as the transmission signal, the rate of speed  is 
\begin{equation}
R_S \sim 1
\end{equation}
So, in general, the speed at input of the channel can be kept at the 
output. This performance is very important to apply physical cipher 
to the real network. 
To derive the real rate of quantum enigma cipher in such cases, 
we will need the information theoretical rate theory. 
 It depends on noise, modulation scheme, error correcting code, 
 scheme of mathematical encryption box, and randomization scheme.
However, IEEE standard for the optical network requires error free 
in point to point link, for example, 
\begin{equation}
P_e(Bob) \sim 10^{-9}
\end{equation}
with no time delay. 
It means that the real application of physical cipher to IEEE standard link 
 requires the following rate:
\begin{equation}
R_E = 1
\end{equation}
The quantum enigma cipher will satisfy the requirements described in the 
introduction of this paper and the above requirement. So,
\begin{equation}
R_E(QEC) \sim 1
\end{equation}
though it depends on the implimentation method.

\subsection{Cryptosystem by quantum enigma cipher}
The quantum enigma cipher consists of an integration of mathematical 
encryption box and physical randomization for ciphertext of mathematical 
encryption box [5].
The mathematical encryption box has a secret key of the length $|K_s|$ bits 
and PRNG for expansion of the secret key. The physical encryption box has 
a mechanism to create ciphertext as signal and it has a function to 
induces an error when the eavesdropper receives the ciphertext as signal.
Consequently the different ciphertext sequences are observed 
in the legitimate's receiver and the eavesdropper's receiver, respectively.
 The quantum mechanics plays a fundamental role to create 
 the following situation.
 \begin{equation}
 P_e(Eve) >> P_e(Bob \quad or \quad Alice) \sim 10^{-9}
 \end{equation} 
 
There are many implementation methods to realize quantum enigma cipher. 
The communication performances may especially depend on the structures of
 both mathematical encryption box and physical randomization.
But since we can employ a high power coherent state signals
for the quantum enigma cipher, we have no signal arrival probability 
problem. In addition, it may have error free communication. 
As a result, in general the quantum enigma cipher does not degrade 
 any input rate. Thus, we can simply replace the tranceiver in the 
conventional optical communication infrastructure (fiber or space) 
 by the quantum enigma cipher tranceiver. This is also very important
  in the real world network.
 
Let us denote some examples. 
Schemes [9,10] based on difference of quantum detection performance and 
quantum illumination [11,3] do not degrade the rate, because
 the rate is kept in the physical randomization in the such 
 simplest cascade schemes.
 
However, a scheme by PPM signal structure may have a tradeoff 
for security-rate [12]. 
In the case of quantum noise randomized stream cipher 
 $\alpha/\eta$[13,14] or Y-00 [15,16] , basically the rate  
is not degraded through the encryption process. However, 
when the masking effect to randomize the ciphertext is very small
 in the real setting, we have to employ 
 additional randomization methods or a new technique.
 Although well known randomizations [17] [18] do not degrade the input rate, 
 some methods such as deliberate randomization in the data stream may 
 degrade it.  In this case, we have to design the randomization carefully.

\section{Conclusion}
This note has introduced the communication performance of QKD and 
quantum enigma cipher which is essential 
for application of all quantum cryptography to the real world.
Consequently, it has been explained that the communication performance
 of QKD by single photon has no prospect.
This fact had been confirmed already in 2003 by several governments.
 However, the development of such QKDs was tolerated by such governments 
 under the several reasons. Now time has come to change the direction of the 
  development of a quantum cryptography with a new concept, because 
  the ciber attack against physical layer of networks become a real 
  possibility at present.

\section*{Acknowledgment}
I am grateful to M.Sohma, F.Futami and K.Kato 
for fruitful discussions, and Russian Academy of Science.



\end{document}